# *Network growth models: A behavioural basis for attachment proportional to fitness*

Michael Bell[*], Supun Perera, Mahendrarajah Piraveenan, Michiel Bliemer, Tanya Latty, Chris Reid

University of Sydney
Sydney, NSW 2008
Australia
*Corresponding author: michael.bell@sydney.edu.au

Several growth models have been proposed in the literature for scale-free complex networks, with a range of fitness-based attachment models gaining prominence recently. However, the processes by which such fitness-based attachment behaviour can arise are less well understood, making it difficult to compare the relative merits of such models. This paper analyses an evolutionary mechanism that would give rise to a fitness-based attachment process. In particular, it is proven by analytical and numerical methods that in homogeneous networks, the minimisation of maximum exposure to node unfitness leads to attachment probabilities that are proportional to node fitness. This result is then extended to heterogeneous networks, with supply chain networks being used as an example.

## INTRODUCTION

Network analysis has emerged as an effective way of studying complex and distributed systems, because it often offers a fine balance between simplicity and realism in modelling such systems [1]. Preferential-attachment models, such as the Barabási-Albert model where attachment probabilities are proportional to target node degree, have been the most prominent among them in the past decade [2]. Recently, a number of node fitness-based attachment models [3, 4, 5, 6, 7, 8, 9, 10, 11, 12] have been gaining prominence. It is noted that the concept of node fitness in this context is parallel to the concept of utility, as used in discrete choice models, where utility is a dimensionless measure of attraction which can be expressed as a function of attributes weighted by their relative importance (for further information regarding the role of utility, refer to [13]). Similarly, node fitness is a dimensionless measure of the inherent competitive ability (or the 'attractiveness') that a node has, which influences the rate at which it acquires links from other nodes as the network evolves over time. One such attribute could be node degree, which if used results in the Barabási-Albert model, but other attributes of the node also could be considered individually or collectively.

An important feature of networks that evolve through node-degree and node-fitness based attachment growth models is that they can exhibit scale free topology, which is found in abundance among biological, technical and social networks [6, 14]. It has been widely accepted that the node-fitness based attachment models (such as the Bianconi-Barabási model [4] or the Ghadge model [3]) are more realistic when compared to the classical Barabási-Albert model in capturing the growth process of real world networks [4, 6]. Nevertheless, the behavioural bases from which such fitness-based attachment behaviour would arise, and the evolutionary mechanisms which shape that behaviour, are less clearly understood. In particular, while many papers have offered intuitive and general explanations as to why increased fitness of a node would attract more links [3, 4, 5, 6], the proportionality between fitness and attachment probabilities assumed in these models is yet to be explained by a rigorous evolutionary framework.

A chain is only as strong as its weakest link, goes an old adage. This paper proposes the minimisation of maximum exposure to least fit nodes (or the avoidance of the "weakest links") as a behavioural basis for the fitness-based proportional attachment rule. To prove this hypothesis, we analyse complex networks where each node is assigned a fitness value, and attachments between nodes are based on node probabilities that minimise the maximum exposure to network unfitness, where unfitness of the network is defined as the inverse of the summation of node fitness values. Two cases are considered:

1. *Homogeneous networks*, where all nodes are of the same type so attachment is possible between any pair of nodes. Note that we use the word 'homogeneous' in this sense and not in the sense that the topology is homogeneous (e.g. a lattice) as is sometimes the case in literature [15, 16, 17]. Some real world examples of such networks could be social networks, both online and offline [16, 17, 18], contact networks in epidemiological studies where any individual can come into contact with any other individual [19, 20, 21, 22], or the World Wide Web (WWW) where any website could link to any other website [6, 16], to name a few. In fact, most real world social, biological, and technical networks studied in the literature will fall into this category, where link formation is not constrained by node type [6, 16]. In our behavioural model for homogeneous networks, attachments are based on node selection probabilities that minimise maximum exposure to the least fit nodes.

2. *Heterogeneous networks*, where nodes are distinguished by type, and only links between certain types are feasible. Networks characterised as bi-partite or k-partite graphs in the literature [16, 23] are heterogeneous networks in this sense. Some real world examples of such heterogeneous networks include collaboration networks where the actors and collaborative activities are both represented as nodes [24], genes and conditions both represented as nodes in microarray data, customers and items both represented as nodes in collaborative filtering [23], pollination networks where pollinators feed only on a subset of available flowers [25], or contact networks in illnesses that spread not from person to person directly but through a vector, such as dengue or malaria [26, 27].

The constraints on attachment in heterogeneous networks can take many forms, and these constraints are context-dependent. Some of these constraints are trivial. Therefore, rather than attempting to analyse all sub-types of heterogeneous networks that can possibly arise from such constraints, we choose here to focus on one interesting sub-type: tiered networks, where each tier contains nodes of a certain type unique to that tier, and from any given tier, only connections to/from the tier 'below' or 'above' are feasible. Hierarchical and multi-layer (tiered) networks have been much studied in recent years, both in terms of topology and in the context of networked games [28, 29, 30, 31, 32, 33], and networks of this type are frequently used to represent supply chains, where four tiers are commonly modelled, namely suppliers, manufacturers, distributers and retailers [34, 35]. Other real world examples for such tiered networks may include tiered bipartite graphs used in gene regulatory network analysis, where for example, a top layer of unobserved regulators can connect with a bottom layer of observed mRNA expression values [36], or food webs where each trophic level (producer, primary consumers, secondary consumers, etc) serves as a tier and connections are only permitted between adjacent tiers [37]. In tiered networks, feasible sets of nodes take the form of paths, so the problem is to find path selection probabilities that minimise maximum exposure to least fit nodes, and then base attachments on the path selection probabilities.

The two cases are distinguished by the node types available and the constraints on link formation, not by resultant topology. Indeed, fitness-based proportional attachment models have been used to model each case, which is why we study them, and each case can result in scale-free topologies with power-law degree distributions [3, 4, 5, 6, 15, 16, 17, 18, 23, 25, 34, 35]. In particular, tiered networks such as supply chain networks are often scale-free with power-law degree distributions [38, 39], even though the tiered visualisation often obscures this.

For each case (homogeneous networks and heterogeneous networks), we formulate a linear program, which *minimises the maximum exposure to network unfitness.* By solving the linear programs, we prove (analytically and numerically) that, *when network unfitness is defined as inverse of the summation of node fitness, minimising maximum expected network unfitness leads to a node attachment probability that is proportional to the node fitness.* Rearrangement of the corresponding Lagrangian equations reveals an equivalent mini-max problem [40], which has a useful game theoretic interpretation, as we will show[41]. An iterative solution process, which mimics a mixed-strategy, non-cooperative, two-player, zero-sum game, is then presented for each case, with convergence to equilibrium proven. For the case of path selection, the inequality constraints in the linear program generate dual variables, which convert the mini-max problem into a shortest path problem, thereby avoiding full set or path enumeration. We then implement this iterative solution process as a computer program and compute the attachment probabilities, which confirm our results obtained by direct solution of the linear program. We use networks of finite size in each numerical example.

The iterative solution process describes an evolutionary mechanism whereby weaker nodes are progressively avoided. Therefore, we demonstrate that such a mechanism could result in attachment probabilities that are proportional to node fitness.

## BACKGROUND

Scale-free networks are ubiquitous [15, 16, 17, 42, 43, 44, 45, 46]. They display power-law degree distributions [2, 17, 42, 47], and are impressively robust to random node failure or damage [48]. However, they are vulnerable to carefully designed targeted attacks [17, 48, 49, 50]. The growth mechanisms responsible for the prevalence of the scale-free networks has been an intensely researched area in the past decade, and the preferential attachment mechanism (the most important element in the Barabási-Albert model) is the most well-known explanation for the prevalence of scale-free networked structures [2, 15, 42, 43]. The preferential attachment mechanism stipulates that the probability of a new node making a link with an existing node is proportional to the number of links (degree) of the existing node. That is, the probability $p_i$ that a new node makes a connection to an existing node $i$ with degree $k_i$ is given by:

$$p_i = \frac{k_i}{\sum_{j \in N} k_j}$$ (Eq. 1)

where *N* is the set of nodes to which the new node could connect. Therefore, this simple degree-based attachment is a rich-get-richer mechanism, where nodes with already high degree are more likely to acquire more links. It is also a mechanism where the degree of a particular node is strongly correlated to its age [2, 15, 42, 47].

In reality, there are factors other than the age and the number of existing connections, which influence the ability of a node to acquire further links. Web pages, companies, and actors all have intrinsic qualities that influence the rate at which they can acquire links [6, 47]. For example, in the World Wide Web, even though Google was a relative late-comer, it quickly overtook other search engines such as Alta Vista and Inktomi in both performance and the number of links, and very quickly became the biggest hub of the World Wide Web [47]. Similarly, in other contexts (for example, the selection of sexual partners, the selection of a firm to acquire supplies from, or the process of deciding to send friend requests in Facebook) it can be seen that late comers can become relatively 'famous' in terms of their connectivity.

In order to explain late-comers acquiring links relatively quickly, a growth model has to take into account the intrinsic property of being desired as a connection by other nodes. In network science this property is called the 'fitness' of the node [3, 4, 5, 7, 8, 9, 10, 29, 47, 51, 52]. The concept of node fitness can be thought of as the amalgamation of all the attributes of a given node that contribute to its propensity to attract links. Indeed, one of these attributes could be the node degree, which would be a dynamic attribute that changes value as the network grows, whereas many other attributes of fitness would be static.

Hence, a number of studies have proposed alternative growth mechanisms based on the notion of node fitness. In many of these studies, preferential attachment is not used at all, and it is argued that the underlying fitness distributions are directly responsible for the emergence of scale-free networks and other well-known topologies in the corresponding systems [7]. A good example of this approach is the study of Caldarelli et al [8] which proposed a good-get-richer mechanism that gives rise to power law degree distributions in networks. In this mechanism, links between two nodes are made with a probability, which is a symmetric function of the fitness of both nodes. Caldarelli et al [8, 9] show that when the underlying node fitness distribution follows a power law, the resultant network will display a power law degree distribution. By invoking Zipf's law [53], they argue that in many real world systems, power law fitness distributions emerge, leading to power law node degree distributions without a preferential attachment mechanism. This observation is presented as an alternative explanation for the prevalence and robustness of power law node degree distributions, implying that preferential attachment models are not the only explanation for power law node degree distributions [7, 10, 12].

Fitness-based proportional attachment models (which use the intrinsic fitness of the nodes as the quantity to compute the preference with) can be thought of as hybrids of the two classes of growth models mentioned above. They are hybrids in the sense that they take 'preferential attachment' from models such as the Barabási-Albert model, and 'fitness' from models such as the Caldarelli model. Although preferential attachment is not needed for the emergence of scale-free networks, these hybrids are attractive because they can be used to model the emergence of scale-free networks even when the underlying fitness distribution is not power law [47]. A prominent example of such a hybrid model is the Bianconi-Barabási model [4, 54], which represents the attachment probabilities as:

$$p_i = \frac{k_i \phi_i}{\sum_{j \in N} k_j \phi_j}$$ (Eq. 2)

where $k_i$ represents the degree of node *i* and $\phi_i$ represents its fitness. As can be seen from the above formulation, between two nodes *i* and *j* with the same fitness $(\phi_i = \phi_j)$, the one with the higher node degree will have the higher probability of selection. Conversely, between two nodes *i* and *j* with the same degree $(k_i = k_j)$ the node with the higher fitness will be selected with a higher probability. In contrast to the Barabási-Albert model, it is possible in the Bianconi-Barabási model for a relative newcomer to overtake an older node in terms of number of links [47]. Empirical studies, such as that by Kong et al., [55] support the Bianconi-Barabási model.

Another hybrid model which uses both fitness distributions and proportional attachment is Lognormal Fitness Attachment (LNFA) presented by Ghadge et al [3, 5]. In the LNFA model, the fitness $\phi_i$, which represents the propensity of node *i* to attract links, is formed from the product of relevant attributes:

$$\phi_i = \prod_{k \in L} \varphi_{ik}$$ (Eq. 3)

When the number of attributes are sufficiently large and statistically independent, it is shown that node fitness $\phi_i$ will be lognormally distributed, regardless of the type of distribution of the attributes [3, 5]. The probability of connecting a new node *j* to an existing node *i* is assumed to be proportional to its fitness as follows:

$$p_i = \frac{\phi_i}{\sum_{j \in N} \phi_j} \qquad \textbf{(Eq. 4)}$$

In contrast to the Bianconi-Barabási model, the LNFA model does not explicitly consider the node degree, and the propensity of a node to attract links is purely dependent on the node fitness. Therefore in LNFA, despite being in the network for a short period of time, a new node which has a large fitness can make itself a preferential choice for other new nodes entering the network. As explained above, this is a reasonable representation of certain real life network growth processes. The model includes a tunable parameter, the shape parameter of the lognormal distribution, which can be varied to generate a wide spectrum of networks corresponding to different real world contexts.

Fitness-based proportional attachment models described above have relied on the intuition that the higher the fitness, the more likely a newcomer is to link to it. However, this intuition about a positive correlation does not necessarily imply the *proportional* attachment formula used in these models, which is only one of several ways in which relatively high-fitness nodes can have high connection probabilities. As such, in this paper, we offer a plausible behavioural mechanism for fitness-based proportional attachment, by proposing the *minimisation of maximum network unfitness* as the underlying process.

## RESULTS

The Table 1 defines the variables and sets used in model derivation.

*Table 1: variables and sets used in model derivation*

| Variables: | |
|---|---|
| $p_j$ | Probability of using node $j$, also the attachment or connection probability |
| $\phi_j$ | Fitness of node $j$ |
| $U_j$ | Inverse fitness (or unfitness) of node $j$ ($U_j = 1/\phi_j$) |
| $q_j$ | Dual variable for node $j \in N$ |
| $h_r$ | Probability that path $r \in R$ is selected |
| $\delta_{jkr}$ | 1 if node $j \in N_k$ lies on path $r \in R$ and 0 otherwise (node $j$ may lie in multiple paths) |
| $\phi_{jk}$ | Fitness of node $j \in N_k$ |
| $U_{jk}$ | Unfitness of node $j \in N_k$ |
| $V$ | Network unfitness |
| Sets: | |
| $N$ | Set of nodes |
| $R$ | Set of all feasible paths (or supply chains), whereby all paths share a common origin and a common destination |
| $K$ | Set of tiers |
| $N_k$ | Set of nodes in tier $k, k \in K$ |

**Case 1 - Homogeneous networks:** Previous work on fitness-based attachment models has generally focused on homogeneous networks where any node can attach to any other node [3, 4]. Although many networks are not homogeneous in this sense, we deal with this case first. We show here that the principle of minimising maximum exposure to unfitness results in the fitness-based proportional attachment model, offering a plausible behavioural basis for attachment *proportional* to fitness.

We formulate a linear program, which chooses attachment probabilities that minimise maximum expected network unfitness, in the following way (here network unfitness is defined as the inverse of the sum of node fitness values):

$$P_0 : \min_p V \text{ subject to}$$
$$V \geq p_j U_j, \forall j \in N$$
$$\sum_{j \in N} p_j = 1$$
$$p_j \geq 0, \forall j \in N$$

The solution to this problem is:

$$V^* = p_1^* U_1 = p_2^* U_2 = \ldots = p_{|N|}^* U_{|N|}$$

where the asterisk denotes a solution value. This in turn implies attachment probabilities proportional to node fitness:

$$p_1^* = V^* / U_1 = f_1 / \sum_{j \in N} f_j$$
$$p_2^* = V^* / U_2 = f_2 / \sum_{j \in N} f_j$$
*etc.*
$$p_{|N|}^* = V^* / U_{|N|} = f_{|N|} / \sum_{j \in N} f_j$$

and minmax expected network unfitness:

$$V^* = \frac{1}{\sum_{j \in N} \phi_j}$$
**(Eq. 5)**

See Methods for an analytical proof. To demonstrate the validity of the above solution (given by Eq. 5) numerically, we begin first with smaller systems that can be solved by the Excel Solver. Thus, we generated a set of nodes (with 12 nodes in the set) with fitness values sampled from a log-normal fitness distribution with scale parameter $\mu = 0$ and shape parameter $\sigma = 1.0$, since it has been argued in the literature that fitness distributions are typically log-normal [3] (Log-normal distributions are characterised by scale and shape parameters; see [3] for a discussion on their relationship with the mean and standard deviation of the distribution). For this set of nodes, we solved the above linear program using the Excel Solver. The results are presented in Fig. 1a, which also shows attachment probabilities calculated as being proportional to fitness. The results match exactly. This result does not depend on the nature or standard deviation of the underlying fitness distribution, and in fact we have verified that any fitness distribution will yield the same result.

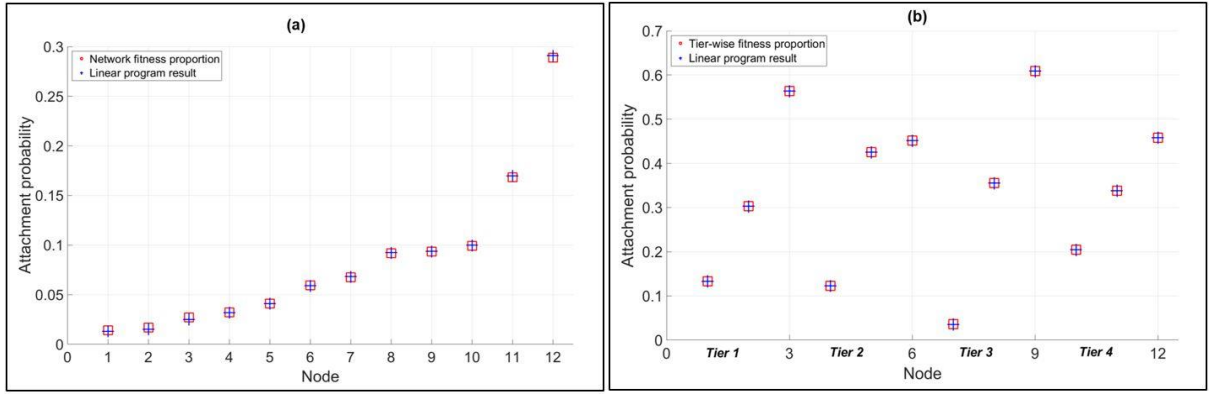

*Figure 1: Linear program solutions for the minimisation of maximum expected unfitness of network (defined as inverse of the summation of node fitness), superimposed on attachment probabilities calculated as proportional to node fitness. Nodes are ordered according to node fitness. (a) homogeneous nodes: the underlying fitness distribution is a log-normal one with shape parameter $\sigma = 1.0$ (b) tiered nodes: the underlying fitness distributions are log-normal ones with shape parameters $\sigma_1 = 3, \sigma_2 = 1, \sigma_3 = 1, \sigma_4 = 0.1$ respectively for the four tiers. In both cases, the linear program solution matches exactly with the attachment probabilities derived from the proportional fitness rule.*

To explore the behavioural basis for attachment proportional to node fitness further (and to devise an algorithm which can easily handle larger systems), consider the following Lagrangian equation for $P_0$:

$$L_{p,q,\lambda} = V + \sum_{j \in N} q_j (p_j U_j - V) + \lambda (\sum_{j \in N} p_j - 1)$$

When this is minimised with respect to $p_j, j \in N$ and maximised with respect to dual variables $q_j, j \in N$ and $\lambda$ the Lagrangian equation has the value of the objective function at the solution and the second and third terms are zero. Rearrangement of the Lagrangian equation gives:

$$L_{p,q,\lambda} = \sum_{j \in N} p_j U_j q_j + V(1 - \sum_{j \in N} q_j) + \lambda (\sum_{j \in N} p_j - 1)$$

Hence $P_0$ is equivalent to:

$P_1 : \max_q \min_p \sum_{j \in N} p_j U_j q_j$ subject to

$\sum_{j \in N} p_j = 1$

$\sum_{j \in N} q_j = 1$

$p_j \geq 0, \forall j \in N$

$q_j \geq 0, \forall j \in N$

As shown in Methods, the simple averaging algorithm $A_0$, presented in Table 2, solves $P_1$ (and therefore $P_0$).

*Table 2: Algorithm $A_0$: Method of Successive Averages for homogeneous networks*

| Step 0: Initialisation | $q_j^* \leftarrow 1/|N|, \forall j \in N; m \leftarrow 1$ |
|---|---|
| Step 1: Find node with lowest weighted unfitness | $p_j \leftarrow 0, \forall j \in H;$ <br> $j^* = \arg\min_{j \in N} U_j q_j^*, p_{j^*} \leftarrow 1$ |
| Step 2: Update node probability | $p_j^* \leftarrow \frac{1}{m} p_j + (1 - \frac{1}{m}) p_j^*, \forall j \in N$ |
| Step 3: Find node with highest expected unfitness | $q_j \leftarrow 0, \forall j \in N;$ <br> $j^* = \arg\max_{j \in N} p_j^* U_j, q_{j^*} \leftarrow 1$ |
| Step 4: Update dual variables | $q_j^* \leftarrow \frac{1}{m} q_j + (1 - \frac{1}{m}) q_j^*, \forall j \in N$ |
| Step 5: Repeat | $m \leftarrow m+1$; return to Step 1 until sufficient convergence is achieved |

Algorithm $A_0$ describes a mixed-strategy, two-player, non-cooperative, zero-sum game whereby one player (call it the system) chooses the node ($j^*$) with the lowest $U_j q_j, \forall j \in N$ (Step 1) and $p_{j^*}$ is increased as a consequence (Step 2). At the same time the other player (call it the demon) chooses the node ($i^*$) with the highest $p_i U_i, \forall i \in N$ and $q_{i^*}$ is increased as a consequence (Step 3). This is an attacker-defender model, where the defender (the system) is choosing a node for attachment subject to the history of unfitness, while the attacker (the demon) is choosing a node to impose unfitness subject to the history of attachment. Hence $\{p_j^*, j \in N\}$ describes the attachment probability while $\{q_j^*, j \in N\}$ describes the unfitness incidence probability.

The demon represents a force or a threat compelling the system to decrease its exposure to less fit nodes. This force (demon) is measured by $\{q_j^*, j \in N\}$. Where $q_j^* = 0$, the demon is exerting no force (note that by the properties of dual variables if $V^* > p_j^* U_j$ then $q_j^* = 0$).

Depending on the context, the demon could be malicious software or Trojans operating in the World Wide Web, diseases that change the topology of intra-cellular networks and metabolic maps [56], or the 'family ties' influence or the so-called *Keiretsu* [57] in supply chains which results in less-fit firms being chosen. The strategy of minimising maximum exposure to network unfitness therefore describes a well-known risk averse strategy.

Algorithm $A_0$ was implemented as a computer program, using a set of 100 homogenous nodes and a log-normal fitness distribution with $\sigma = 1$ (indeed, a technical advantage of such an iterative algorithm is that it could easily solve larger systems which simple linear program implementations such as Excel Solver may struggle with). The results are shown in Fig. 2a. Again, it could be seen that the attachment probabilities generated by the algorithm match exactly with those generated by a proportional attachment model. Fig 2b shows the convergence of solution for two nodes (as examples), and it can be seen that the solutions converge to the exact values predicted by proportional attachment based on fitness. The convergence of Algorithm $A_0$ is proven in Methods.

We have conducted similar experiments with varying system sizes, and varying node fitness distributions. In all instances, we get attachment probabilities, which exactly match those calculated by fitness-based proportional attachment, though the number of iterations for convergence depends on the size of the system.

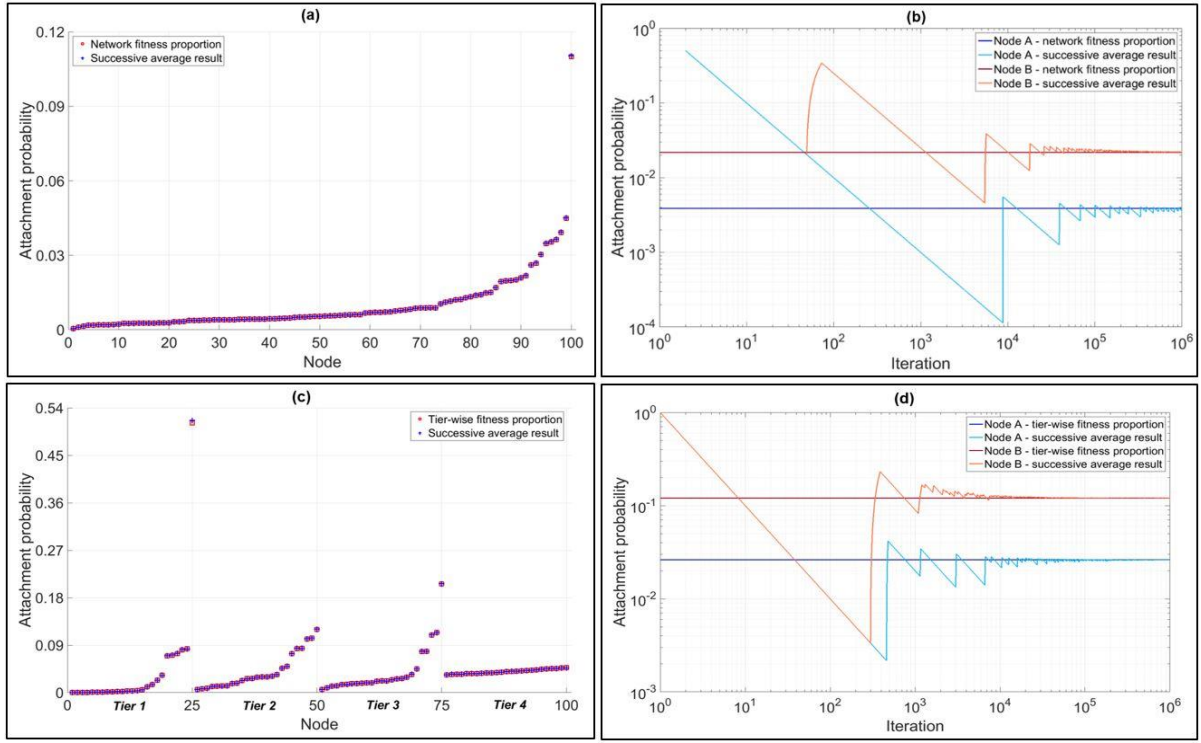

*Figure 2: Convergence and results of iterative algorithms $A_0$ and $A_1$. Nodes are ordered according to fitness. (a) $A_0$ (homogeneous case) - results (b) $A_0$ - Two examples of convergence (c) $A_1$ (tiered case) - results (d) $A_1$ - Two examples of convergence. It can be seen that in both homogeneous and tiered cases, the iterative algorithms eventually converge to solutions that exactly match those calculated by proportional attachment based on node fitness.*

**Case 2 - Heterogeneous networks:** In some networks, nodes are not homogeneous, so connections can only be made between certain types of nodes. Link formation in heterogeneous networks can be subject to different constraints according to the context. Some of these are trivial. For example, in bipartite networks, links are usually made between nodes of the 'opposing' types [58]. In some poly-partite networks too, links can be made between any pair of nodes as long as they do not belong to the same type: for instance, in mentoring programs, people who have disparate levels of seniority and/or skill are usually matched. A first year undergraduate could be mentored by a second, third or fourth year student but not by another freshman. In other contexts, a feasible set must exist before links can be made within that set. For example, if a social network is constructed among individuals who play the game of cricket, each set must have exactly eleven players including batsmen, bowlers and wicket-keepers before they can constitute a team, though once constituted, social links can be made between any two individuals of such a team. Therefore, attachment behaviours prevalent in such heterogeneous networks are also diverse and context-specific.

Rather than making a general study, we here choose to focus on an interesting special case - the tiered network. In particular, we choose to study supply chain networks (SCNs), which are frequently described by such tiered networks, where there are several types of nodes, and only adjacent tiers can be linked. Supply chain networks provide one of the richest examples for a heterogeneous network, because they have nodes of several types, and they possess a hierarchical structure, where any firm (node) can only connect to firms in the tier above or below. Supply chain networks have been studied extensively, and preferential attachment mechanisms have been used to model their growth [34, 35, 38, 39, 57]. Typically, SCNs have four tiers: suppliers, manufacturers, distributors and retailers [34, 35, 38]. Fig. 3 shows a network representing a tiered SCN.

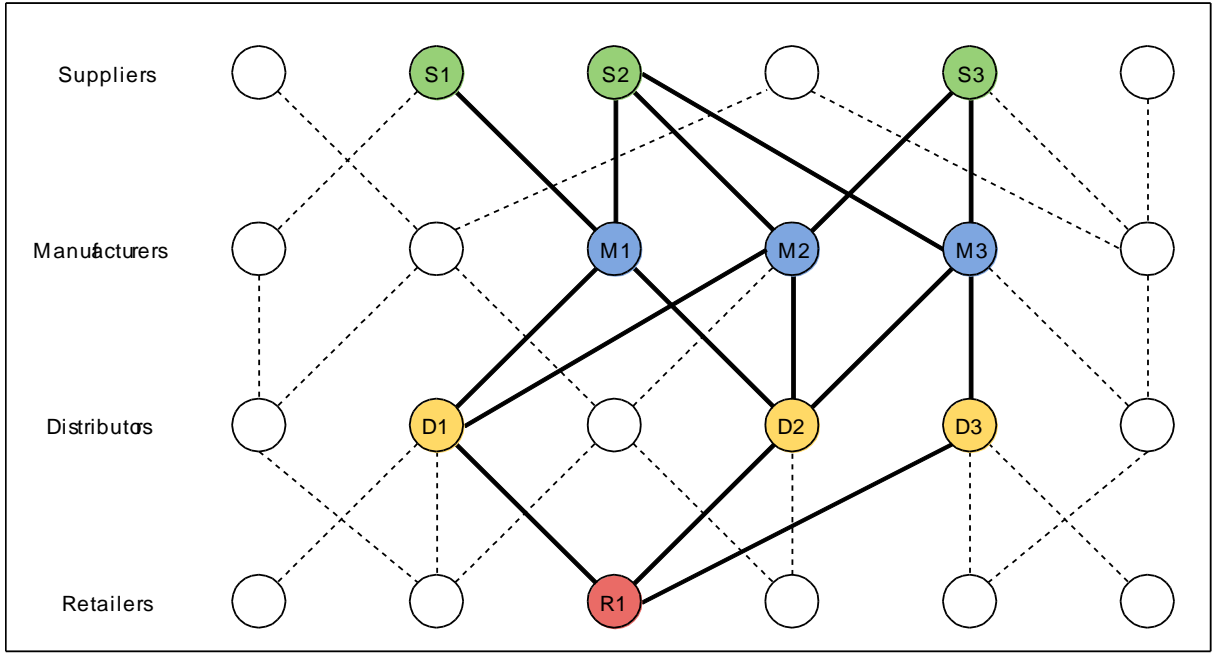

*Figure 3: Conceptual diagram of a SCN, highlighting the supply chain for one retailer R1 (red node). Note that there are four tiers, and each plausible path (supply chain) has to connect a node from each tier in a particular order.*

Let us therefore consider the case where each firm minimises its maximum exposure to unfitness in the tier immediately above. This reflects reality in many supply chains, as firms in one tier often do not have visibility beyond the tier immediately upstream [34, 35]. This leads to the following linear program:

$$P_2 : \min_p \sum_{k \in K} V_k \text{ subject to}$$

$$V_k \geq p_{jk} U_{jk}, \forall j \in N_k, \forall k \in K$$

$$\sum_{j \in N_k} p_{jk} = 1, \forall k \in K$$

where the index *k* refers to a particular tier in the SCN. Since every feasible path (equivalent to feasible sets in a more general heterogeneous network scenario) must pass through each tier, the node selection probabilities must sum to one for each tier. The path (set) selection probabilities may thus be replaced by node selection probabilities so problem $P_2$ reduces to $|K|$ independent problems and can be solved as such:

$$p_{ik}^* = V_k^* / U_{ik} = \phi_{ik} / \sum_{j \in N} \phi_{jk}, i \in N_k$$

with minmax expected tier unfitness for each tier *k* given by:

$$V_k^* = \frac{1}{\sum_{j \in N} \phi_{jk}} \quad \textbf{(Eq. 6)}$$

The linear program formulated above was solved using Excel Solver, and as an example, we used a set of 12 nodes, which belonged to four types (tiers) in equal number. The fitness distributions were again log-normal, with scale parameter $\mu = 0$ and shape parameters $\sigma_1 = 3, \sigma_2 = 1, \sigma_3 = 1, \sigma_4 = 0.1$ for suppliers, manufacturers, distributors and retailers respectively. The shape parameters used in each tier, as noted above, represent a SCN with diminishing oligopoly conditions from upstream (suppliers) to downstream (retailers). At the retailer level, the differences in node fitness across the tier are small, so the retailers of this product can be assumed to be in almost perfect competition. An example of such a SCN is for computer spare parts, where valuable minerals such as Coltan [59] are sourced from suppliers who tend to be in an oligopoly market in unique geographical locations, such as the Democratic Republic of Congo. Subsequently, the manufacturers and distributors of these unique and specialised parts are also likely to be in oligopoly markets, although to a lesser extent when compared to the raw material suppliers. Finally, the retailers of these computer parts will be almost homogeneous with no significant differentiation in 'fitness'.

The solution for this particular example is presented in Fig. 1b. It could be seen that the node selection probabilities computed by fitness-based proportional attachment match exactly the solution of the linear program again. Again, we have verified that this result does not depend on the fitness distributions of tiers, and any fitness distribution for each tier will yield the same result, though in the example presented we have used 'realistic' fitness distributions as described above.

Now, in order to arrive at an iterative algorithm which leads to the above solution (and is able to handle larger systems), let us note that $P_2$ may be re-expressed as the following max-min problem:

$$P_3 : \max_q \min_p \sum_{j \in N, k \in K} p_{jk} U_{jk} q_{jk} \text{ subject to}$$

$$\sum_{j \in N_k} p_{jk} = 1, \forall k \in K$$

$$\sum_{j \in N_k} q_{jk} = 1, \forall k \in K$$

This leads to the following Lagrangian equation:

$$L_{p,q,\lambda} = \sum_{j \in N, k \in K} p_{jk} U_{jk} q_{jk} + \sum_{k \in K} V_k (1 - \sum_{j \in N_k} q_{jk}) + \sum_{k \in K} \lambda_k (1 - \sum_{j \in N_k} p_{jk})$$

The structure of $P_3$ reveals that the problem is equivalent to a two-player uncooperative mixed strategy game, where one player (call it a scheduler) seeks to minimise network unfitness by choosing a path while tier-specific 'demons' seek to maximise the same metric by imposing unfitness on one node per tier.

The evolutionary algorithm to find the solution requires modification from $A_0$ so that maximum unfitness is minimised within each tier. We now undertake the process of finding the shortest path (which is equivalent to finding the best set in the more general heterogeneous network case). The modified algorithm is given as Algorithm $A_1$ in Table 3.

*Table 3: Algorithm $A_1$ : Method of Successive Averages for tiered networks*

| Step 0: Initialisation | $q_{jk}^* \leftarrow 1/|N_k|, \forall j \in N_k, \forall k \in K; R \leftarrow \varnothing; m \leftarrow 1$ |
|---|---|
| Step 1: Find best path | $h_r \leftarrow 0, \forall r \in R;$ <br> $r^* = \arg\min_{r \in R} \sum_{j \in N_k, k \in K} \delta_{jkr} U_{jk} q_{jk}^*, h_{r^*} \leftarrow 1;$ <br> $R \leftarrow R \cup \{r^*\}$ |
| Step 2: Update node usage | $p_{jk}^* \leftarrow \frac{1}{m} \delta_{jkr^*} h_{r^*} + (1 - \frac{1}{m}) p_{jk}^*, \forall j \in N_k, \forall k \in K$ |
| Step 3: Find the node most exposed to unfitness | $q_{jk} \leftarrow 0, \forall j \in N_k, \forall k \in K;$ <br> $\forall k \in K, j^* = \arg\max_{j \in N_k} p_{jk}^* U_{jk}, q_{j^*k} \leftarrow 1;$ |
| Step 4: Update node duals | $q_{jk}^* \leftarrow \frac{1}{m} q_{jk} + (1 - \frac{1}{m}) q_{jk}^*, \forall j \in N_k, \forall k \in K$ |
| Step 5: Repeat | $m \leftarrow m + 1$ ; return to Step 1 until sufficient convergence achieved |

In Algorithm $A_1$, during each iteration, we find the least unfit path and set the corresponding auxiliary path probability equal to one (implying that all other auxiliary path probabilities are zero). In finding the least unfit path in Step 1, each node unfitness is weighted by the current value of $q_{jk}^*, j \in N_k, k \in K$, which is the relative frequency with which $p_{jk}^* U_{jk}$ is found to be largest in each tier. When $p_{j^*k}^* U_{j^*k}$ is largest in its tier in Step 3, the corresponding auxiliary variable $q_{j^*k}$ is set to one (implying all other auxiliary variables in the tier are zero). This causes the value of $q_{j^*k}^*$ to increase and the value of $q_{jk}^*, j \neq j^*$ to decrease, as shown in Step 4. The Algorithm tells us that the final value of $q_{jk}^*, j \in N_k, k \in K$ depends on the effect which a change in node unfitness $U_{jk}$ would have on tier unfitness $V_k^*$, and this depends on how critical node $j$ is to the tier.

Again, to show that the iterative algorithm $A_1$ converges to attachment probabilities which could be calculated by fitness-based proportional attachment, we implemented $A_1$ as a computer program. Here we show a sample solution by considering a set of 100 nodes, which belonged to four types (tiers) in equal number. The nodes had log-normal fitness distributions with zero scale parameter and shape parameters $\sigma_1 = 3, \sigma_2 = 1, \sigma_3 = 1, \sigma_4 = 0.1$ for suppliers, manufacturers, distributors and retailers respectively. Convergence was achieved, and the results are shown in Fig. 2c, where we may see that the solution matches exactly with attachment probabilities computed by fitness-based proportional attachment. Fig 2d shows the convergence of solution for two nodes (as examples), and it can be seen that the solutions converge to the exact values predicted by proportional attachment based on fitness. We repeated this experiment for different system sizes and found that while the number of iterations needed depended on the system size, the results were otherwise identical (converging always to values predicted by fitness-based proportions).

The supplementary materials contain additional insights about the experiments. Fig 4 and Fig 5 in the supplementary materials illustrate the visualisations obtained for 1,000 node homogeneous and tiered network scenarios, respectively.

## DISCUSSION

In summary, we have proposed the minimisation of maximum exposure to network unfitness as an evolutionary mechanism that results in the proportional fitness-based attachment rule used in network growth models. We have studied homogeneous networks, and an especially interesting and illuminating subset of heterogeneous networks, namely tiered networks. We have proven, both analytically and numerically, that for both homogeneous and tiered networks, the abovementioned mechanism leads to attachment probabilities which are proportional to (node) fitness. For both categories of network, we have also developed iterative algorithms with proven convergence, which offer a behavioural template possibly mimicking the cognitive decision making process that results in fitness-based attachment. Our results shed light on the validity and interpretation of a range of fitness-based growth models which are used to study complex network evolution in various contexts.

It is worth to point out here that our results justify both the Bianconi-Barabási and the Ghadge models of proportional attachment. The Bianconi-Barabási model [4] calculates attachment probabilities as proportional to the product of node degree and fitness, whereas the Ghadge model [3] computes them as proportional to just the fitness. The mechanism of minimising the exposure to maximum unfitness that we propose here, at first glance, seems to support the Ghadge model, since it returns attachment probabilities proportional to node fitness. However, in reality, players (nodes) may view other players (nodes) who have a high number of links as more desirable: that is, they could appear to have relatively higher 'fitness'. Therefore, the Ghadge and Bianconi-Barabási models could be reconciled by making the fitness score endogenous to the model, and making node degree one factor in the fitness calculation. Once this is done, the mechanism we have proposed is equally relevant to explain both Bianconi-Barabási and Ghadge models.

We have further shown that for heterogeneous or tiered networks, an evolutionary mechanism, whereby least unfit sets or paths are iteratively generated, can lead to the minimization of maximum exposure to unfitness. We have noted that this mechanism has the added advantage that it avoids the full enumeration of sets or paths. In the case of tiered networks, such as supply chain networks, the mechanism explains how nodes (for e.g., business firms) can make informed choices (e.g., in terms of selecting suppliers) even without fully analysing the entire set of feasible paths.

## METHODS

**Complementary slackness conditions:** Consider the Lagrangian equation for the minimum maximum exposure to unfitness problem $P_0$ in the homogeneous network case:

$$L_{h,q,V,\lambda} = V + \sum_{j \in N} q_j (p_j U_j - V) + \lambda (\sum_{j \in N} p_j - 1)$$

As a result of the complementary slackness conditions, at the solution (denoted by *)

$$q_j^* > 0 \Rightarrow p_j^* U_j = \max_{i \in N} p_i^* U_i = V^*, j \in N$$

and

$$p_j^* U_j < p_i^* U_i \Rightarrow q_j^* = 0, i, j \in N$$

**Non-negativity of primal and dual variables in homogeneous networks:** All primal variables are positive when all values of node fitness are positive.

*Lemma 1:* $p_j > 0, \forall j \in N$ in the homogeneous case for $f_j > 0, \forall j \in N$

Proof 1: $p_j^* = V^* / U_j = f_j / \sum_{j \in N} f_j > 0, \forall j \in N$ since $f_j = 1/U_j$.

**Equivalence of primal and dual variables in homogeneous networks for positive fitness values:** In the case of homogeneous networks, primal variables are equal to corresponding dual variables when fitness values are positive.

*Lemma 2:* $p_j^* = q_j^*, \forall j \in N$ and $V^* = \lambda^*$ for homogeneous networks when node fitness values are positive.

*Proof 2:* Let us recall that the principle of choosing attachment probability to minimize maximum expected network unfitness gives rise to the following linear program:

$P_0 : \min_p V$ subject to

$V \geq p_j U_j, \forall j \in N$

$\sum_{j \in N} p_j = 1$

$p_j \geq 0, \forall j \in N$

Consider the following Lagrangian equation for $P_0$ (note that in Lagrangian formulation, $\lambda$ could be introduced with either a positive or negative sign):

$$L_{p,q,V,\lambda} = V - \sum_{j \in N} q_j (V - p_j U_j) + \lambda (1 - \sum_{j \in N} p_j)$$

Rearranging the above, we get;

$$L_{p,q,V,\lambda} = \lambda - \sum_{j \in N} p_j (\lambda - U_j q_j) + V(1 - \sum_{j \in N} q_j)$$

Comparing the second Lagrangian equation (the Lagrangian for the dual problem) with the first (the Lagrangian for the primal problem) we note that at the solution $V$ is equal to $\lambda$ because the second and third terms are zero at the solution and furthermore $q_j$ is equal to $p_j$ by complementary slackness because $p_j > 0$ when all values of node fitness are positive (see Lemma 1). Hence at the solution,

$$V^* = I^*$$
$$p_j^* = q_j^*$$

**Convergence of the Method of Successive Averages:** Algorithm $A_0$ is presented in Table 2.

*Lemma 3*: Algorithm $A_0$ converges.

*Proof 3*: Step 1 finds a descent direction for $V$ given $\{q_j^*, j \in N\}$ with respect to $\{p_j^*, j \in N\}$ and Step 2 moves the node probabilities in this direction. Step 3 finds an ascent direction for $V$ given $\{p_j^*, j \in N\}$ with respect to $\{q_j^*, j \in N\}$ and Step 4 moves the node dual variables in this direction. While the step sizes get smaller as iterations progress, the sum of step sizes has no upper limit, so the solution can always be reached.

A similar proof of convergence can be constructed for algorithm $A_1$.

ACKNOWLEDGEMENTS

The research underpinning this paper has been funded by the Australian Research Council (ARC) under grant DP140103643.

AUTHOR CONTRIBUTIONS

M.Bell and SP conceptualised the research. M.Bell made the initial findings, devised the evolutionary algorithms, and wrote the manuscript. SP wrote the software and provided the visualisations and numerical examples. MP reviewed the literature, designed the software, and wrote the manuscript. TL, M.Bliemer and CR helped revise the manuscript. All authors contributed to research design, analysis of results, and fine-tuning of ideas.

ADDITIONAL INFORMATION

The authors declare no competing financial interests.